\documentstyle[prl,aps]{revtex}
\draft
\input{psfig.sty}
\begin{document}
\twocolumn[\hsize\textwidth\columnwidth\hsize\csname@twocolumnfalse\endcsname
\title{Crystallization Kinetics in the Swift--Hohenberg Model}
\author{Igor S. Aranson$^1$, Boris A. Malomed$^2$, Len M.~Pismen$^3$, and
Lev S.~Tsimring$^4$}
\address{$^{1}$Argonne National Laboratory, 9700
South Cass Avenue, Argonne, IL 60439\\
$^2$ Department of  Interdisciplinary Studies, Faculty of Engineering,
Tel-Aviv University, Tel-Aviv 69978, Israel\\
$^3$Department of Chemical Engineering, Technion, 32000 Haifa, Israel\\
$^4$ Institute for Nonlinear Science, University of California at
San Diego, La Jolla, CA 92093-0402}
\date{\today}

\maketitle

\begin{abstract}
It is shown numerically and analytically that the front propagation process
in the framework of the Swift-Hohenberg model is determined by 
periodic 
nucleation events triggered by the explosive growth of the localized
zero-eigenvalue mode of the corresponding linear problem. We derive the evolution
equation for this mode using asymptotic analysis, and evaluate
the time interval between nucleation events, and hence the front speed.
In the presence of noise, we derive the velocity exponent of ``thermally activated'' front propagation (creep) beyond the pinning threshold. 
\end{abstract}

\narrowtext
\vskip1pc]


One of fundamental models of pattern formation far from equilibrium
is the Swift-Hohenberg (SH) equation
\cite{SH}. Despite relative simplicity of this model, and a serious
limitation related
to the fact that it is of the gradient type, it gives rise to a remarkably
large
variety of solutions. The SH equation has been intensively studied in the past
as a paradigm model of pattern formation in large aspect ratio systems. 
\cite{SH,GC}. More recent computations brought attention
to the propagation of fronts between uniform stationary states of this equation
and coarsening \cite{jap}. The computations have also demonstrated
formation of stationary solitons, i.e. stable localized objects in the form
of a domain of one phase sandwiched inside another phase \cite{jap}. The
interest was supported by applications of the SH model to marginally
unstable optical parametric oscillators (OPO) \cite{tl}, and more numerical
computations, leading to similar conclusions, were carried out in this
context  \cite{stl}. The SH model has also served as a convenient testing
tool for the problem of pattern propagation into an unstable trivial state
\cite{dee,saa}. It is the purpose of this Letter to elucidate another
aspect of pattern formation that can be modeled by the SH equation:
coexistence between a pattern and a uniform state and stick-and-slip motion
of interphase boundary which can be thought of as a particular kind of
crystallization or melting.

\begin{figure}
\centerline{\psfig{figure=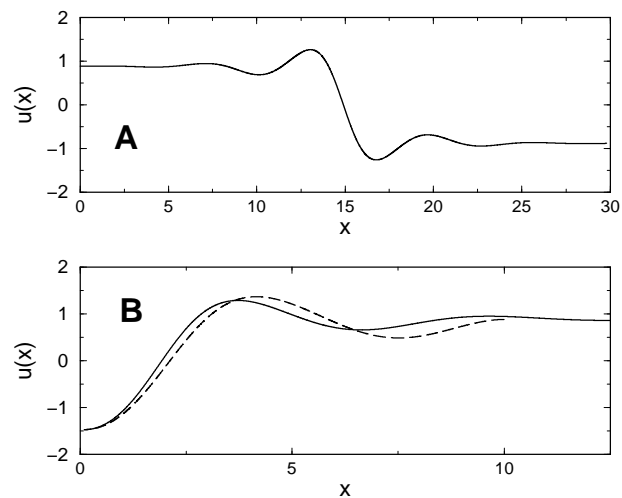,width=3.2in}}
\caption{Stationary solutions to Eq. (\protect \ref{SH}) for $\epsilon=1.78$.
(a) kink solution connecting $u=\pm \sqrt{\epsilon-1}$; (b) 1D localized
solution  (solid line) and 
radially-symmetric solution to  Eq. (\protect \ref{SH}) (dashed line). }
\label{fig1_new}
\end{figure}

We shall write the basic equation in the form
\begin{equation}
u_t=-(1+\nabla^2)^2 u +\epsilon u-u^3.
\label{SH} \end{equation}
At $\epsilon>0$ the trivial state $u=0$ undergoes supercritical stationary
bifurcation
leading to a small-amplitude pattern with unit wavenumber $k$. The band of
unstable
wavenumbers widens with growing $\epsilon$, until it reaches the limiting
value $k=0$,
which signals the appearance of a pair of nontrivial uniform
states, $u=\pm\sqrt{\epsilon-1}$. The two symmetric states are stable to
infinitesimal perturbations at $\epsilon>\frac{3}{2}$. 
At still higher values of $\epsilon$, a
variety of
metastable states become possible: (1) a kink separating the two symmetric
nontrivial
uniform states (Fig.~\ref{fig1_new}a); (2) a semi-infinite pattern, coexisting
with either of
the two nontrivial uniform states; (3) a finite patterned
inclusion,
sandwiched, either symmetrically or antisymmetrically, 
between semi-infinite domains occupied by 
nontrivial uniform states \cite{brand}; (4) a
soliton  (Fig.~\ref{fig1_new}b)
in the form of an island where one of the uniform states is
approached, immersed in the infinite region occupied by the alternative
state; the basic
soliton ``atom''  can be viewed as  a single period of the
pattern immersed in a uniform state.

The energy of the uniform state is higher than that of the
regular pattern with the optimal wavenumber
at $\epsilon<6.287$ \cite{our}. In spite of the
difference between the
energies of the uniform and patterned state, the interface
between them
remains immobile at moderate and large values $\epsilon$, and a multiplicity
of localized states is linearly stable. 
The pattern propagation into a metastable
uniform state or the reverse "melting" process at $\epsilon \gg 1$ is
impeded by the {\it self-induced pinning} attributed
to the oscillatory character of the asymptotic perturbations of the uniform
state, characterized by the complex wavenumber $k^2=-1 \pm
i\sqrt{2\epsilon-3}$.  As $\epsilon$ decreases, the stationary localized
solutions lose stability and give rise to propagating solutions.

We shall show in this Letter that the front propagation process can be
described in terms of periodic
nucleation events triggered by explosive growth of the localized
zero-eigenvalue mode of the corresponding linear problem. Using asymptotic
analysis, we shall derive the evolution equation for this mode which allows
us to estimate the
time interval between the nucleation events and hence the front speed.
We shall also study the ``thermally activated'' front propagation (creep) at
$\epsilon>\epsilon_c$ and derive the creep velocity exponent.

\begin{figure}
\centerline{\psfig{figure=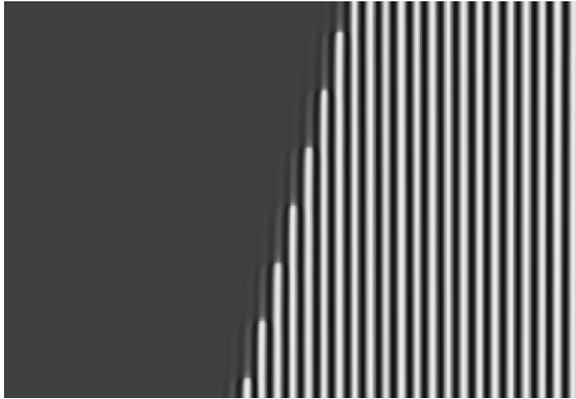,width=3.in}}
\caption{The space-time plot of the switching wave propagation into the stable
uniform phase at $\epsilon=1.757$. The horizontal axis is spatial coordinate,
and the vertical axis is time. The system size is 200 (4096 mesh points),
the time span is 200.}
\label{f1}
\end{figure}

\begin{figure}
\centerline{\psfig{figure=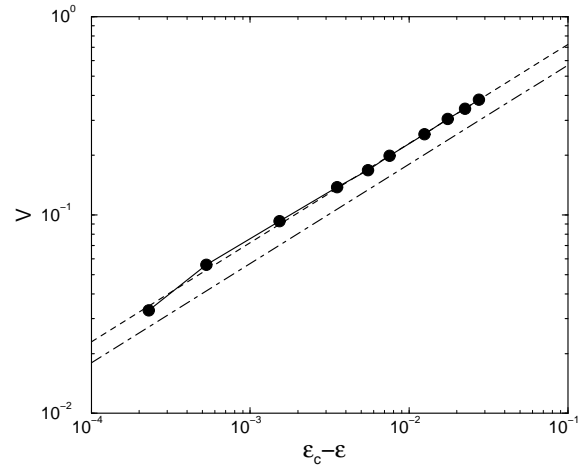,width=3.in}}
\caption{The average front speed as a function of the control parameter $\epsilon$
near the critical point $\epsilon_c=1.7574..$. Points represent numerics,
the dashed line
is the best fit $V=2.29(\epsilon_c-\epsilon)^{1/2}$ and the dot-dashed line is the
theoretical prediction $V=1.80(\epsilon_c-\epsilon)^{1/2}$.
}
\label{f2}
\end{figure}

We performed numerical study of Eq.(\ref{SH}) in one spatial dimension at
various values of $\epsilon$.
We found that the depinning transition for the single soliton occurs at
$\epsilon \approx 1.74$.
The depinning threshold increases with the size
of the patterned cluster, rapidly converging the limiting
value $\epsilon=\epsilon_c \approx 1.7574...$ for the semi-infinite pattern.
Figure \ref{f1} shows propagation of the semi-infinite pattern into
a uniform stable state at $\epsilon=1.75$. The front propagation takes
the form of well separated in time periodic nucleation events of  new ``atoms''
of the ``crystalline'' state
at the front. Between successive nucleation events, the solution remains close to the
stationary semi-infinite pattern found at $\epsilon_c$. This process
resembles crystallization in equilibrium solids, with the important
distinction that the new ``atoms'' are created directly from the metastable
``vacuum'' state. The time between consecutive nucleation events diverges
as $\epsilon$ approaches the pinning threshold \cite{mitkov}. Figure \ref{f2} presents the
average front speed as a function of $\epsilon-\epsilon_c$. This function
can be
fitted by $V=V_0 \sqrt{\epsilon_c-\epsilon}$, with $V_0
=2.292$.


\begin{figure}
\centerline{\psfig{figure=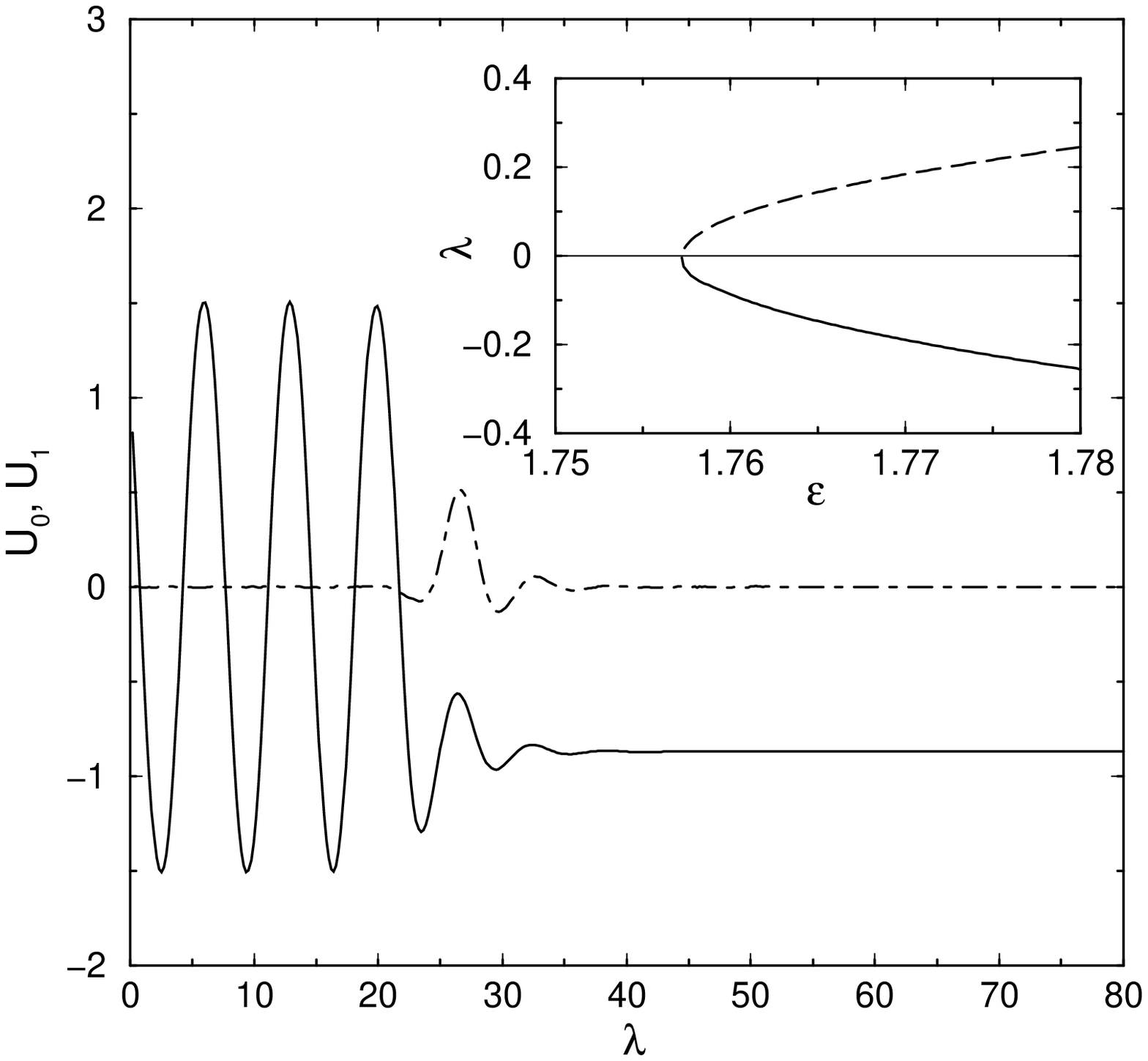,width=3.in}}
\caption{The stationary structure $U_0(x)$ (solid line) and the corresponding localized
zero mode $U_1(x)$ (dashed line) at $\epsilon=\epsilon_c$. 
Inset: eigenvalues of the stable and
unstable
localized modes of the stationary front solutions at $\epsilon>\epsilon_c$.
}
\label{f3}
\end{figure}

At $\delta=\epsilon -\epsilon_c>0$,
the immobile front solution $u_0(x,\epsilon)$ is
linearly stable. Numerical stability analysis shows that
there exists an exponentially decaying mode $u_1(x,\epsilon)$
localized at the front with negative
eigenvalue $\lambda_s$ which approaches zero as  $\delta^{1/2}$.
In addition, there is also an unstable front solution and a corresponding
mode with positive
eigenvalue $\lambda_u$ which also approaches zero as $\delta^{1/2}$.
At the pinning threshold, these two solutions collide
and disappear via a saddle-node bifurcation.
The stationary front solution
$U_0(x)\equiv u_0(x,\epsilon_c)$ and the corresponding mode structure
$U_1(x)\equiv u_1(x,\epsilon_c)$ are shown in Fig.\ref{f3}. In the inset,
we show
the stable and unstable eigenvalues $\lambda_{s,u}$ as functions of
$\epsilon$.
At $\epsilon<\epsilon_c$, the front solution becomes non-stationary.
Nonetheless, at $|\delta| \ll 1$, the numerical study suggests that the
solution remains close to the stationary front solution $U_0(x)$ all the
time except short intervals when a new roll nucleates.

The process of front propagation can be analyzed within the framework of
perturbation theory near the critical value $\epsilon=\epsilon_c$.
At $|\delta|\ll 1$, we can present the front solution
in the form
$$u(x,t)=U_0(x)+|\delta|^{1/2}u_{1}(x,t),$$
where
$U_0(x)$ is the stationary front solution at $\delta=0$ and
$|\delta|^{1/2}u_1(x,t)$ is
a small correction. This solution is uniformly valid  at small positive
$\delta$; it is also valid during quasi-stationary phases (away from nucleation
events)
for small negative $\delta$.
Plugging this ansatz in Eq.(\ref{SH}), we obtain
\begin{equation}
\partial_t u_1=L[U_0]u_1 -|\delta|^{1/2}\left[ 3U_0u_1^2
-\mbox{sign}(\delta) U_0 \right] +...,
\label{mode}
\end{equation}
where $L[U_0]\equiv\epsilon_c-3U_0^2-(1+\nabla^2)^2$ is the linearized SH
operator at $\epsilon_c$.
In the lowest order in $\delta$, Eq.(\ref{mode}) yields the linear equation
$\partial_t u_1=L[U_0]u_1$.
This equation is always satisfied by the stationary translational mode
$ U_{0}'(x)$. In addition, the
linearized operator $L[U_0]$ has a localized neutral
eigenmode $U_1$ (found numerically above, see Figure \ref{f3}).
Since all other eigenmodes have negative eigenvalues, the evolution of the
system
close to the bifurcation point can be reduced to single-mode
dynamics \cite{note}.
Therefore in the lowest order we can write $u_1(x,t)= a(t) U_1(x)$, where
$a(t)$ is the amplitude of the zero mode.

Close to the bifurcation threshold $\partial_t u_1 = \dot a U_1$ can be
treated as
perturbation.  Therefore, in the second order we derive
\begin{equation}
L[U_0]u_2 = \dot a U_1+ |\delta|^{1/2}\left[3a^2 U_0 U_1^2
 - \mbox{sign}(\delta)U_0 \right].
\label{mode2}
\end{equation}
Eq.~(\ref{mode2}) has a bounded solution if its r.h.s. is orthogonal to
the zero mode $U_1$ of the operator $L$. This results in the solvability
condition for
the amplitude $a$:

\begin{equation}
\alpha\, \dot a = |\delta|^{1/2}\left[\mbox{sign}(\delta)\beta - \gamma a^2
\right],
\label{solv}
\end{equation}
where
$$
\alpha = \int_{-\infty}^{\infty} U_1^2 dx, \;\;
\beta =  \int_{-\infty}^{\infty} U_1 U_0 dx, \;\;
 \gamma = 3 \int_{-\infty}^{\infty} U_1^3 U_0 dx.
$$

At $\delta>0$, $a(t)$ reaches a stationary amplitude
$a_0=(\beta/\gamma)^{1/2}$. This value corresponds to the difference
between the
stable front solutions at $\epsilon$ and $\epsilon_c$. At small {\em negative}
$\delta$, Eq.(\ref{solv}) describes {\em explosive} growth of $a$, which
passes from
$-\infty$ to $\infty$ in
a finite time $\tau_e=\pi\alpha/(|\delta |\beta\gamma)^{1/2}$. This explosion
time gives
an upper bound for the period between the nucleation events, after which
the whole process repeats. The front speed is found as $V=\Lambda/\tau_e$,
where $\Lambda$ is the
asymptotic spatial period of the pattern selected by the process of roll
nucleation, and $\tau_e$ is the time interval between nucleation events.
Our calculations give the value: $V= 1.8 |\delta|^{1/2}$.
This scaling is in a good  qualitative agreement with the results of the
numerical simulations, see Fig.\ref{f2}. However, the prefactor $1.8$
is noticeably lower (about 25\%) than the corresponding value
$V_0=2.29$ obtained by numerical
simulation of Eq.~(\ref{SH}).

Let us discuss a possible reason for the discrepancy. Our numerical
simulations show that the nucleation events produce slowly-decaying
distortions  behind the moving front. These distortions
may effectively ``provoke" a
consequent nucleation event by creating an initial perturbation of
the zero mode $U_1(x)$. It will lead to an increase of the front velocity.
Although we have evidence for the importance of this effect, a systematic
treatment of this process is very complicated and goes beyond
perturbation theory.

\begin{figure}
\centerline{\psfig{figure=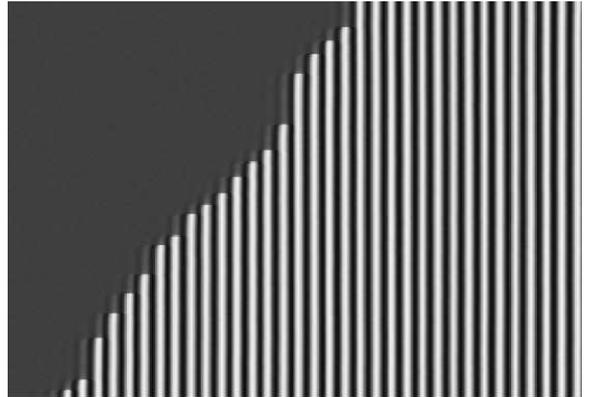,width=3.in}}
\caption{The space-time plot of the switching wave propagation
triggered by white noise with the temperature $T=0.00042$ at
$\epsilon=1.758>\epsilon_c$. The horizontal axis is spatial coordinate,
and the vertical axis is time. The system size is 200 (4096 mesh points), the
time span is 1000.}
\label{f4}
\end{figure}

\paragraph*{Effect of noise.} Let us consider the effect of weak additive
noise
$\eta(x,t)$ in the r.h.s. of Eq. (\ref{SH}). For simplicity we assume that
$\eta$ is delta-correlated noise with the intensity (temperature) $T$:
$$\langle \eta(x,t) \eta(x^\prime , t^\prime) \rangle = T \delta(x-x^\prime)
\delta(t -t^\prime).$$
In the presence of noise, there is no sharp threshold for the onset of
motion at $\epsilon<\epsilon_c$. Instead, thermally-activated motion
(creep) occurs at $\epsilon>\epsilon_c$ (see Fig.~\ref{f4}).
In this case the average creep velocity is determined by the intensity of noise. For
$\epsilon<\epsilon_c$ the noise will slightly increase the speed of the
front. In contrast to the deterministic motion, the intervals between
consecutive nucleation events are random.

In order to estimate the effect of the noise at $\delta>0$, we will treat $\eta$ as small
perturbation. In this analysis, we shall not introduce scaling explicitly,
since the scaling of noise that is comparable by its effect with
deterministic perturbations cannot be determined {\it a priori}.
Following the lines of the above analysis, we project
noise at the zero mode to obtain the solvability condition
\begin{equation}
\alpha\, \dot a = \delta^{1/2}(\beta - \gamma a^2)+ \delta^{-1/2}\tilde
\eta(t) ,
\label{solv1}
\end{equation}
where $\tilde \eta(t) = \int_{-\infty}^{\infty} \eta(x,t) U_1(x) dx$.
For $T=0$,  Eq. (\ref{solv1}) has a stable fixed point
$a_s=\sqrt{\beta/\gamma}$ and
an unstable one $a_u=-\sqrt{\beta/\gamma}$. At $a<a_u$, one has explosive
growth of the solution
to Eq. (\ref{solv1}), while $a\to a_s$ at $a>a_u$. Thus, we have to
estimate the probability $P$ for the amplitude of the zero mode $a$ to be
smaller than $a_u$. This quantity can
be derived from the corresponding Fokker-Planck equation for the
probability density $p(a,t)$ \cite{vK}:
\begin{equation}
p_t = -\frac{\delta^{1/2}}{\alpha} \frac{\partial}{\partial a}\left[(
\beta-\gamma a^2)p \right]+
\frac{T}{2 \alpha \delta } \partial_a^2 p ,
\label{fpe}
\end{equation}
where we used $\langle \tilde \eta^2\rangle = \alpha T $. This is the
standard Kramers problem.
The stationary probability $p(a)$ is given by
\begin{equation}
p \sim \exp\left[\frac{2 \delta^{3/2} (\beta a -\gamma a^3/3)}{T}\right] .
\label{p}
\end{equation}
The probability $P$ is given by the integral
$P = \int _{-\infty}^{a_u} p(x) dx$. For $ T \ll \delta^{3/2}$, we can
use the saddle-point method, which gives the following result:
\begin{equation}
P(a<a_u) \sim \exp\left[ -\frac{4 \delta^{3/2}  \beta^{3/2}}{3\gamma^{1/2} T
}\right] =
\exp\left[ -\frac{0.57  \delta^{3/2} }{T }\right] .
\end{equation}
Since the time between the nucleation events $\tau_n \sim 1/P$, we find
that the velocity of the front
in the stable region is given by $v \sim 1/\tau_n \sim
\exp\left[ -0.57  \delta^{3/2} /T \right]$.

At very large $\epsilon >\epsilon_{0}=6.287$,
the flat state $u=\pm \sqrt{\epsilon -1}$ has lower energy than the
periodic state.  Nevertheless, the uniform state does invade periodic
state at any $\epsilon$ because  of the strong self-induced pinning.
With large-amplitude noise there will be some probability for the
flat state to
propagate towards periodic state by thermally-activated annihilation events
at the edge of the periodic pattern, however t large $\epsilon$ the
probability of annihilation at the edge is of the same order as that
in the bulk of the patterned state. 
Thus, for very large $\epsilon$ and large $T$ we may expect melting of 
the periodic structure both on the edge and in the bulk.

The above results can be trivially
extended to regular two-dimensional periodic structures (rolls)
selected by the SH equation \cite{SH}. More interesting is the behavior of a
2$D$ hexagonal lattice which was predicted by weakly nonlinear
analysis near the bifurcation of nontrivial uniform solutions at
$\epsilon=\frac{3}{2}$ and found numerically \cite{dewit}.  We
anticipate that propagation of the hexagonal structure into the uniform 
state will exhibit 
the same features of stick-and-slip motion as described above, and can be 
studied by similar methods. For SH equation (\ref{SH}) at 
$\epsilon>\frac{3}{2}$, hexagonal lattices coexist with roll patterns which 
have lower energy, so the front propagation may in fact give rise to
rolls. However, the hexagons may become dominant in the modified SH equation 
with an added quadratic nonlinearity near $\epsilon=0$, where the robust 
``crystallization'' of the hexagonal lattice is expected. The work on
this subject is now in progress.

The authors thank the Max-Planck-Institut f\"ur Physik komplexer Systeme,
Dresden, Germany for hospitality during the Workshop
on Topological Defects in Non-Equilibrium Systems and Condensed Matter.
ISA and LST  acknowledge support from the U.S. DOE under grants
No.    W-31-109-ENG-38, DE-FG03-95ER14516, DE-FG03-96ER14592 and
NSF, STCS \#DMR91-20000. LMP acknowledges the support by the Fund for
Promotion of Research at  the Technion and by the Minerva
 Center for Nonlinear Physics of Complex Systems.


\begin{thebibliography}{80}
\bibitem{our}
A close value of $\epsilon_{0}=6.3$ was obtained in Ref. \protect\cite{jap}
using relaxation method. We obtained a more accurate estimate of
$\epsilon_{0}$ and the corresponding optimal wavenumber $k_0=0.933$ by
matching-shooting method for stationary Eq. (\protect\ref{SH}). The above
values were derived  by comparing the energy of the uniform state
$E_0=-\frac{1}{2}(\epsilon-1)^2$
with the energy of the periodic state with the optimal wavenumber.
\bibitem{note} The translational mode $U'_0$ of course also has a zero
eigenvalue.
However, its mobility vanishes, since the friction factor
$\int_{-\infty}^{\infty}U_0^{\prime2}dx$ is infinite, and
therefore this mode can be excluded.
\end{thebibliography}
\end{document}